\documentclass[aps,prd,twocolumn,amsmath,showpacs,superscriptaddress,floatfix,nofootinbib,
nopreprintnumbers]{revtex4-1}
\usepackage{verbatim}
\usepackage[T1]{fontenc}
\usepackage[utf8]{inputenc}
\usepackage[american]{babel}
\usepackage{epsfig}
\usepackage{xspace}
\usepackage{graphicx}
\usepackage{booktabs}
\usepackage{multirow}
\usepackage{dcolumn}
\usepackage{amsmath}
\usepackage{mathtools}
\usepackage{amsfonts}
\usepackage{amssymb}
\usepackage{epstopdf}
\usepackage{siunitx}
\usepackage{braket}
\usepackage{enumitem}
\usepackage{soul}
\usepackage{subfiles}
\usepackage[table]{xcolor}
\usepackage{color}
\usepackage{transparent}
\usepackage{pifont}
\usepackage[normalem]{ulem}
\usepackage{bm}
\usepackage{comment}
\usepackage{subfiles}
\usepackage{orcidlink}
\usepackage{hyperref}
\hypersetup{
    colorlinks=true, 
    linkcolor=blue, 
    citecolor=magenta}

\newcommand{\mdark}{$m_{\rm dark}$\xspace}

\newcommand{\dchigamma}{$\Delta\chi^2$\xspace}

\begin{document}
\title{Shedding light on dark matter spikes through refractive neutrino masses} 
\author{Federica Pompa \orcidlink{0000-0002-9591-8361}}
\email{federica.pompa@subatech.in2p3.fr}
\affiliation{SUBATECH, IMT Atlantique, CNRS/IN2P3, Nantes Université, Nantes 44307, France\\}
\author{Manibrata Sen \orcidlink{0000-0001-7948-4332}}
\email{manibrata@iitb.ac.in}
\affiliation{Indian Institute of Technology Bombay, Bombay Powai, Mumbai 400076, India\\}
%
%
\begin{abstract}
The origin of neutrino mass remains an open question in particle physics. One intriguing possibility is that neutrinos are massless in vacuum but acquire an effective refractive mass through interactions with ultralight dark matter during propagation. We investigate the capability of the upcoming Deep Underground Neutrino Experiment (DUNE) to probe such refractive masses using the time-of-flight delays of neutrinos from a galactic core-collapse supernova. Our analysis shows that DUNE can set competitive bounds on the refractive neutrino mass, with sensitivity significantly enhanced if neutrinos traverse a dark matter density spike near the Galactic Center. In particular, we quantify how the presence of a spike modifies the projected limits, demonstrating that supernova neutrino observations at DUNE provide a powerful and novel avenue to test both the nature of neutrino masses and the distribution of dark matter in the innermost regions of the Milky Way. 
\end{abstract}
\maketitle
%
\noindent\textbf{\emph{Introduction -- }}The existence of neutrino oscillations implies that neutrinos have mass.  However, after decades of theoretical and experimental effort, the origin of neutrino masses is still unknown. 
The non-zero nature of neutrino masses presents evidence of the existence of physics beyond the Standard Model (SM). Hence, measurements of neutrino mass provide a powerful window into the underlying new physics, from seesaw mechanisms to interactions with exotic backgrounds such as dark matter (DM).

Neutrino masses can be generated through the usual spontaneous symmetry breaking, like the rest of the SM fermions~\cite{Englert:1964et,Higgs:1964pj}, or they can be sourced through some new interactions~\cite{Davoudiasl:2018hjw, Choi:2019zxy,Choi:2020ydp,Chun:2021ief,Hui:2016ltb,Smirnov:2021zgn,Sen:2023uga,Sen:2024pgb}. We refer to the former as the \emph{vacuum neutrino mass} ($m_{\rm vac}$), while we call the latter \emph{refractive neutrino mass} ($m_{\rm dark}$). A popular example of refractive neutrino mass was considered in~\cite{Sen:2023uga}, where the authors explored the contribution to neutrino masses induced by coherent forward scattering with the surrounding DM background, similar to the Mikheyev-Smirnov-Wolfenstein (MSW) effect~\cite{Mikheev:1986wj,Wolfenstein:1977ue}. This mass is inherently different from the vacuum mass, and its magnitude depends on the integrated DM density encountered along the neutrino trajectory. 

Ref.~\cite{Sen:2023uga} argued that, since neutrino oscillations are sensitive to the mass-squared (and not to the mass directly), any contribution to the Hamiltonian of the form ${(\rm const/}E)$ can reproduce the observed oscillation parameters.
However, if the DM background maintains its coherence, it can induce time-modulations in the neutrino oscillation parameters, which can be constrained from existing data~\cite{Berlin:2016woy,Brdar:2017kbt,Capozzi:2018bps,Dev:2020kgz,Losada:2021bxx,Huang:2022wmz, Dev:2022bae,Davoudiasl:2023uiq,Martinez-Mirave:2024dmw,Sen:2024pgb,Goertz:2024gzw,Sahu:2025vyy}. 
Using a combination of neutrino data, primarily from the KamLAND experiment, and arguing that $\mathcal{O}(1)$ fluctuations of DM can average out the effects of such time modulations, Ref.~\cite{Cheek:2025kks} set bounds that disfavor this mechanism as the dominant source of neutrino mass.
However, it remains to be seen if such DM field can maintain its coherence over cosmological timescales due to virialization in the halo. Furthermore, the treatment of DM fluctuations through simplified statistical models may not fully capture the complexity of realistic galactic substructure, leaving room for alternative interpretations.

The detection of neutrinos from a galactic core-collapse supernova (SN) provides a unique opportunity to probe the absolute neutrino mass through precise time-of-flight measurements. This method has been already employed with neutrinos from SN1987A \cite{Kamiokande-II:1987idp,Bionta:1987qt,Alekseev:1988gp,Alekseev:1987ej}, from which a $95\%$ confidence level (C.L.) upper limit of $m_{\rm vac}<5.8$ eV has been derived \cite{Pagliaroli:2010ik,Loredo:2001rx}.
The same method has also been used to evaluate the neutrino mass sensitivity for various upcoming neutrino experiments~\cite{Pompa:2022cxc,Pompa:2023yzg,Denton:2024mlb}.
While traditional analyses constrain $m_{\rm vac}$, they do not account for the possibility that neutrinos may acquire additional in-medium contributions while propagating through high-density DM regions. 

A recent study~\cite{Ge:2024ftz} proposed using the time-of-flight delay of SN neutrinos to distinguish between $m_{\rm vac}$ and $m_{\rm dark}$.
This approach highlights how future galactic SN observations can probe the origin of neutrino masses while simultaneously providing information on the DM distribution in the Milky Way.

The time delay scales with the square of the effective neutrino mass $m_\nu = \{m_{\rm vac}, m_{\rm dark}\}$, and can be significantly enhanced when the neutrino trajectory passes through an increased DM density, also called a DM spike. Such DM spikes can form, for example, around a massive black hole (BH) due to the adiabatic growth of the BH itself within a pre-existing DM halo~\cite{Gondolo:1999ef, Balaji:2023hmy}, or due to accretion by primordial BHs formed during the radiation-dominated era~\cite{Carr:2020mqm}.
The increase in time delay becomes most pronounced in the case of a galactic SN occurring beyond the Galactic Center (GC), with neutrinos traversing the high-density DM region on their way to Earth. While previous studies have explored the connection of refractive masses with SN neutrino time-of-flight measurements, they have generally assumed smooth halo DM profiles and neglected the effects of possible DM spikes. In contrast, our analysis explicitly connects the structure of DM in the inner galaxy to observable neutrino timing signatures at DUNE, offering a novel and complementary probe of DM spikes. \textcolor{black}{This observable probes the time-averaged quadratic refractive mass term, which remains unconstrained by the oscillation bounds discussed in Ref.~\cite{Cheek:2025kks}.}
The potential to test the existence of such spikes through alternative astrophysical channels has also been highlighted in recent works~\cite{Tiwari:2025qqx,Feng:2024obn,Fujiwara:2023lsv, Ferrer:2022kei, Cline:2023tkp,DelaTorreLuque:2024wfz, Dev:2025tdv}.

\begin{figure}[h]
    \centering
\includegraphics[width=\columnwidth]{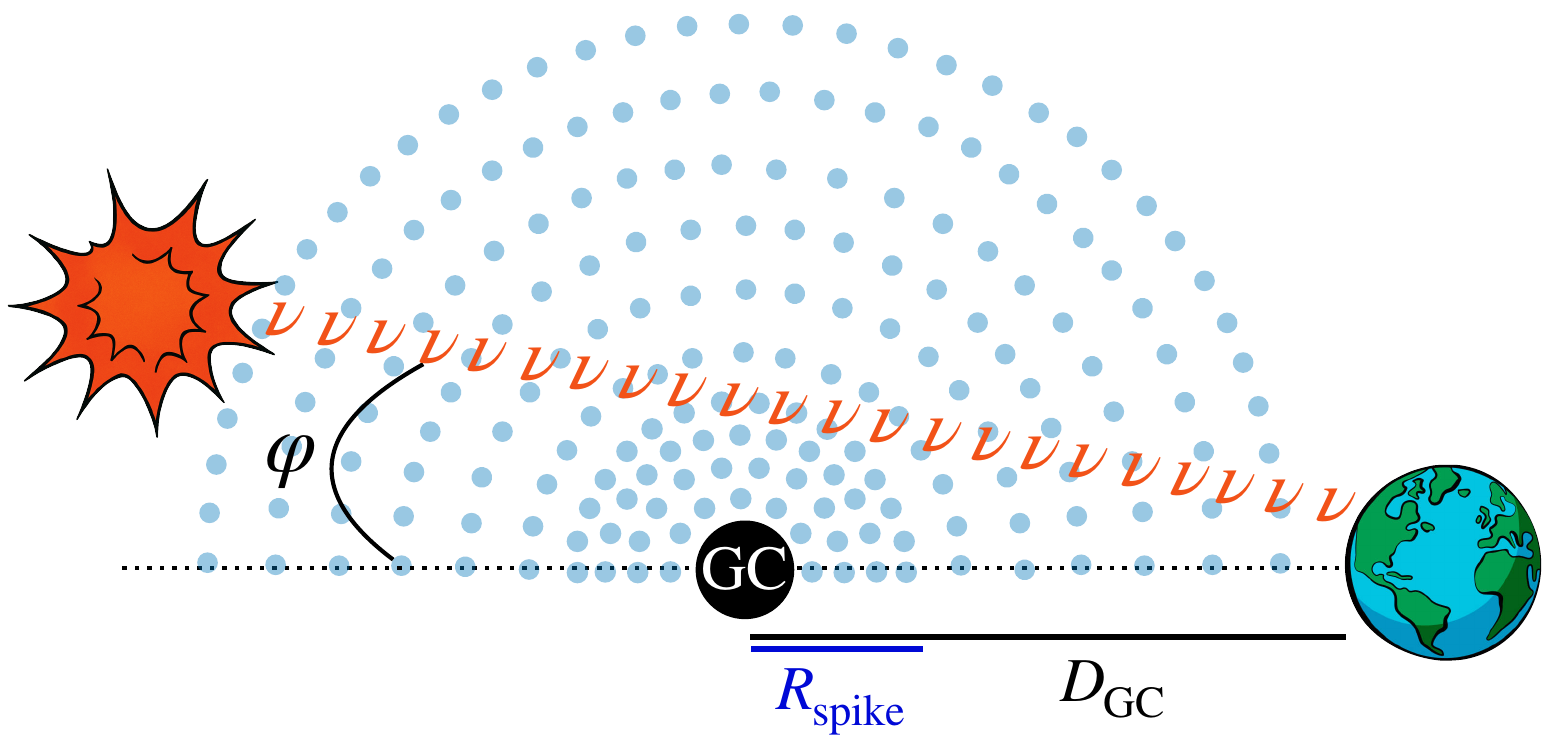}
    \caption{Image depicting neutrinos from a SN traveling through the DM spike to reach the Earth. The GC is located at a distance of $D_{\rm GC} = 8$ kpc from the Earth, and the DM spike region around it extends up to a radius $R_{\rm spike}$. The $\varphi$ angle defines the angular distance of the SN neutrino flux from the DM spike, defined following \cite{Ge:2024ftz}.}
    \label{fig:cartoon_spike}
\end{figure}

In this work, we explore the sensitivity of DUNE to refractive neutrino masses in scenarios where DM spikes are present near the GC (see Fig.~\ref{fig:cartoon_spike} for a representative case). 
We show that the line-of-sight (LOS) integral of the DM density leads to a measurable increase in the time-of-flight delay of SN  neutrino by upto an order of magnitude compared to the case of a smooth halo. Our results highlight how future SN neutrino observations can serve as a novel probe of both the nature of neutrino mass and the DM distribution in the innermost galactic regions.

\medskip

\noindent\textbf{\emph{Neutrino potential and refractive mass --  }}In this framework, neutrinos acquire an effective refractive mass through coherent forward scattering on a background of ultralight dark matter (ULDM). 
The interaction is mediated by a light fermionic field, producing a potential that depends on the 
number density of ULDM particles and the cosmological redshift, and is given by $ V = m^2_{\rm dark}(y - \epsilon)/(2 E_R (y^2 - 1))$~\cite{Sen:2023uga},
where $y \equiv E_\nu/E_R$ and the resonance energy $E_R = m_f^2/(2m_\phi)$, where $m_f$ is 
the mass of the fermionic mediator and $m_\phi$ is the DM mass. The DM asymmetry is captured by $\epsilon$. The refractive mass is given by $m^2_{\rm dark} \equiv (g^2\, \rho_\phi)/m_\phi^2 $,
where $\rho_\phi$ is the DM density and $g$ is the coupling between neutrinos and DM.

For neutrino energies above the resonance ($E_\nu \gg E_R$), the refractive potential becomes $V \approx m^2_{\rm dark}/2 E$, thus
reproducing the behavior of the conventional mass-squared differences measured 
in oscillation experiments. Below the resonance, the refractive mass decreases with neutrino 
energy, and its dependence is shaped by the charge asymmetry of the DM.
While $m^2_{\rm dark}$ reproduces the properties of a vacuum mass 
above the resonance, it declines rapidly with energy below the resonance and cannot be used in the 
same way as a usual mass term. This scaling implies that during the epoch of structure formation, 
refractive masses were too small to affect clustering, effectively rendering relic neutrinos massless. 
Such behavior provides a pathway to reconcile neutrino oscillation measurements with the strong 
cosmological bounds on the sum of neutrino masses~\cite{Sen:2024pgb,DESI:2025zgx}.

\noindent\textbf{\emph{Time delay within the DM spike -- }}The idea of constraining the neutrino mass with SN neutrinos relies on measuring the time-of-flight delay \cite{Zatsepin:1968ktq} experienced by a neutrino of vacuum mass $m_{\rm vac}$ and energy $E_\nu$, traveling a distance $D$ before detection, $\Delta t_{\rm vac} = D\,m_{\rm vac}^2/(2\,E_{\nu}^2 )$\,. \textcolor{black}{Here $m_{\rm vac}$ denotes the effective neutrino mass relevant for time-of-flight measurements, analogous to the kinematic mass probed in beta decay experiments such as KATRIN. 
This standard approximation in SN neutrino propagation studies (see Refs.~\cite{Pompa:2022cxc,Ge:2024ftz}) holds since the arrival-time spread among mass eigenstates is far below experimental timing resolutions.}%

This expression applies to neutrinos propagating without interacting with DM. However, when neutrinos with refractive masses traverse DM rich regions, the time delay formalism must be modified in order to include the effects due to neutrino propagation in DM halos~\cite{Ge:2024ftz}. Furthermore, the presence of a supermassive black hole (SMBH), such as Sgr~A$^\star$ at the GC, is 
expected to significantly enhance the surrounding DM density, creating a steep DM spike~\cite{Gondolo:1999ef}. Over cosmic time, the intense gravitational pull of the 
SMBH can draw DM inward, compressing the halo profile and producing densities far higher than those 
predicted by standard halo models. Such spikes are of particular interest because they can 
substantially amplify indirect signals of DM, including neutrino refractive effects relevant to 
this work.

However, the modeling of DM spikes remains highly uncertain. The inner kiloparsec of the Milky Way 
is dominated by baryons, making it difficult to disentangle their influence on the DM profile. 
Stellar heating, feedback processes, and past merger events can soften or even partially erase 
spikes, while DM self-annihilation imposes an upper limit on achievable central densities. Moreover, 
assumptions about whether Sgr~A$^\star$ has remained stationary at the GC and the 
precise stellar distribution nearby further complicate predictions. As a result, while DM spikes 
are theoretically well motivated, their actual structure and density remain open questions that 
must be probed observationally.

One possible way of probing such DM spikes is through the time delay induced on neutrinos propagating through them. 
To describe this scenario, we assume a symmetric DM distribution and neglect annihilation processes. 
The spiked DM density profile is parameterized by the piecewise function~\cite{Balaji:2023hmy}:
\begin{equation}
    \rho_\phi(r, \gamma) = 
\begin{cases}
0 & \quad \text{if } r < 2R_{\rm S}\,\,, \\
\rho_{\text{spike}}(r, \gamma)\,\, & \quad \text{if } 2R_{\rm S} \leq r < R_{\text{spike}}\,\,, \\
\rho_{\text{\tiny NFW}}(r) & \quad \text{if } r \geq R_{\text{spike}}\,\,,
\end{cases}
\end{equation}
where $R_{\rm S} = 2.95 (M_{\rm BH}/M_\odot)$~km is the Schwarzschild radius of the BH 
($M_{\rm BH} = 4.3 \times 10^6 M_\odot$ for SgrA$^\star$ at the GC), $R_{\text{spike}}$ is the radius of extent of the spike,
$\rho_{\text{\tiny NFW}}(r) = \rho_0 / \left[(r/r_s)(1+r/r_s)^2\right]$ is the Navarro--Frenk--White (NFW) 
halo profile, with $\rho_0 = 0.34~\text{GeV}/\text{cm}^3$, and
$\rho_{\text{spike}}(r,\gamma) =
\rho_{\text{\tiny NFW}}(R_{\rm spike})\left(r/R_{\rm spike}\right)^{-\gamma}$
describes the spike contribution, with $\gamma$ controlling its steepness.  

\textcolor{black}{
For ULDM, one must ensure that the spike is stable against wave-like delocalisation. 
The relevant scale is the de~Broglie wavelength
$\lambda_{\rm dB}=h/(m_\phi v)$, where $v$ is the local virial velocity in the spike region. Because $v(r)\sim\sqrt{GM_{\rm BH}/r}$ increases toward the GC, $\lambda_{\rm dB}$ decreases correspondingly. This implies that the condition $\lambda_{\rm dB}\lesssim R_{\rm spike}$ is satisfied for $m_\phi \gtrsim \mathrm{few}\times 10^{-22}\,\mathrm{eV}$, i.e., in the same range where
ULDM halos are known to retain coherence. We therefore work in the region where the spike profile is self-consistent.}

The corresponding time delay can be estimated as
\begin{equation}
    \label{eq:delay-dark}
    \Delta t_{\rm dark}(\varphi,\gamma) = \frac{D}{2}\left(\frac{m_{\rm dark}}{E_{\nu}}\right)^2 \frac{\overline{\rho_\phi}(r_\star,\varphi,\gamma)}{\rho_\phi(r_\odot)}\,,
\end{equation}
where $\rho_\phi(r_\odot) = 0.43$ GeV/cm$^3$ is the local DM energy density in the Solar System, and $ \overline{\rho_\phi}(r_\star, \varphi, \gamma) =  \int_{x_\star}^{x_{\odot}} \rho_\phi(r(x, \varphi, \gamma))~dx/D$
is the average DM energy density along the neutrino’s LOS. The angle $\varphi$ defines the trajectory orientation relative to the galactic plane, with $\varphi = 0^\circ$ corresponding to 
propagation along the plane and $\varphi = 90^\circ$ perpendicular to it, as defined in Fig.~\ref{fig:cartoon_spike}.

\begin{figure}[!t]
    \centering
\includegraphics[width=\columnwidth]{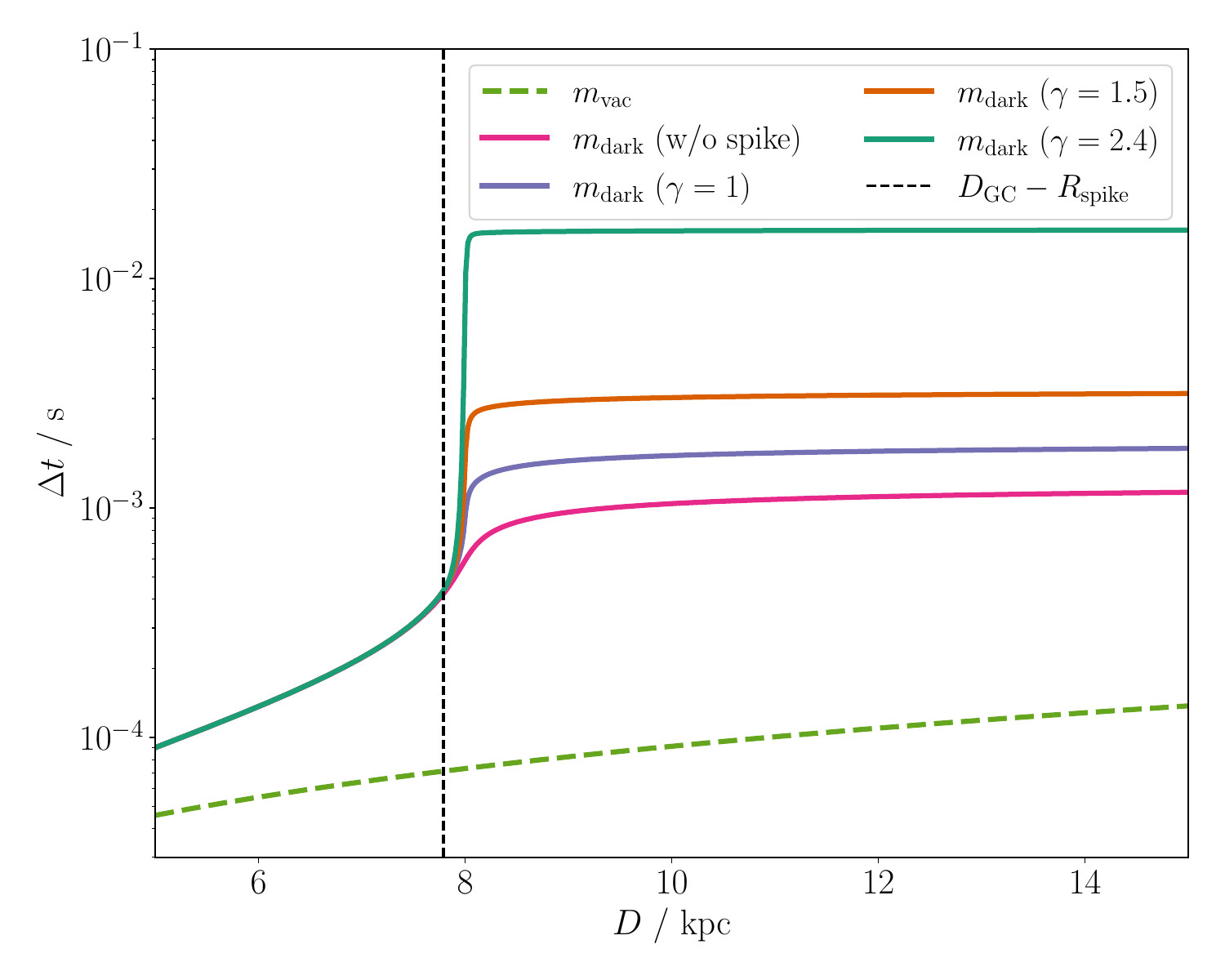}
\centering
    \caption{Time-of-flight delay of SN neutrinos as a function of the SN distance $D$, assuming as a reference a neutrino (vacuum or dark) mass value of $0.2$ eV, and energy of $15$ MeV. For the dark-mass case, results are shown both with and without a DM spike, considering different spike density slopes $\gamma$. The dashed vertical line marks the outer edge of the spike region, extending to $R_{\rm spike} = 0.21$ kpc from the GC (see Fig.~\ref{fig:cartoon_spike} for notation).}
\label{fig:Dt_VS_distance}
\end{figure}

Fig.~\ref{fig:Dt_VS_distance} illustrates how the neutrino time delay varies with the SN 
distance and the spike parameter. The enhancement of $\Delta t_{\rm dark}$ for trajectories crossing the DM spike is compared 
with both the vacuum case ($\Delta t_{\rm vac}$, dashed light-green line) and the halo case without a 
spike (solid pink line). The vertical dashed line indicates the spike region near the GC. The behavior of the time delay with distance can be understood as follows. Within the DM halo and prior to encountering the spike, $\Delta t_{\rm dark}$ scales approximately linearly with $D$, as the average DM density along the LOS remains close to the local value $\rho_\phi(r_\odot)$. Once neutrinos cross the spike, the LOS density increases sharply, causing $\Delta t_{\rm dark}$ to level-off at a value set by the angular distance $\varphi$ and the spike normalization $\gamma$.

\noindent\textbf{\emph{Supernova neutrino events in DUNE -- }}Core-collapse SNe release almost all their gravitational binding energy ($\simeq 10^{53}$ ergs) in the form of $\mathcal{O}(10)$ MeV (anti)neutrinos of all flavors.
Neutrino emission happens within a timescale of about $\mathcal{O}(10)$ s, and can be divided into three distinct phases: the initial \emph{neutronization burst}, the \emph{accretion phase}, and the final \emph{cooling phase}. 
For time-of-flight delay measurements, the neutronization burst is of particular importance. Lasting about $25$~ms after core bounce, this phase features a sharply peaked 
flux of electron neutrinos ($\nu_e$) that is weakly dependent on the progenitor star’s 
properties and the SN’s hydrodynamical evolution \cite{Kachelriess:2004ds,Mirizzi:2015eza}.
These features make the SN neutronization phase a strong tool to extract robust neutrino mass bounds via time delay measurements \cite{Pompa:2022cxc}.

The differential neutrino spectra for each neutrino flavor emitted by the SN core bounce, can be described by the quasi-thermal (alpha-fit) parameterization discussed in \cite{Tamborra:2012ac,Keil:2002in,Lang:2016zhv}, which is consistent with detailed numerical simulations. As they propagate outward, neutrinos undergo adiabatic flavor conversions due to coherent 
forward scattering with stellar matter through the MSW effect \cite{Wolfenstein:1977ue, Mikheev:1986wj}. Collective neutrino oscillations are neglected for this study~\cite{Mirizzi:2015eza}, whereas 
Earth matter effects are omitted since their 
impact on neutrino mass sensitivity is negligible~\cite{Pompa:2022cxc}.

The DUNE far detector, employing liquid argon technology, will be uniquely sensitive to the $\nu_e$ 
component from the next galactic SN via the charge-current (CC) channel $\nu_e + {^{40} Ar} \rightarrow e^{-} + {^{40} K^{*}}$~\cite{DUNE:2015lol}. 
The expected event rate can be estimated as $R(t,E) = N_\text{target}~\sigma_{\nu_e\text{CC}}(E)~\epsilon(E)~\Phi_{\nu_e}(t,E)$,
where $N_\text{target}=6.03\times 10^{32}$ gives the number of argon nuclei for a $40$ kton liquid argon fiducial mass, $\sigma_{\nu_e\text{CC}}(E)$ is the $\nu_e$ CC cross-section taken from \texttt{SNOwGLoBES} \cite{snowglobes}, and $\epsilon(E)$ the DUNE reconstruction efficiency with a 5~MeV threshold \cite{DUNE:2020zfm}.

If the neutrinos were massless, then the detection time would read as $t_d=t + D$. 
However, their finite mass induces a time delay $\Delta t$, leading to $t_d=t+D+\Delta t$. Typically, detectors are sensitive to the time difference between the $i$-th event and the first one detected, $\delta t_d^i$, so the common parameter $D$ drops out. Hence, the time of detection for the $i$-th event can be related to the time of emission as $\delta t_d^i= t^i + \Delta t$, where the emission time of the first event is set to zero by default.

\begin{figure}[tb]
    \centering
\includegraphics[width=\columnwidth]{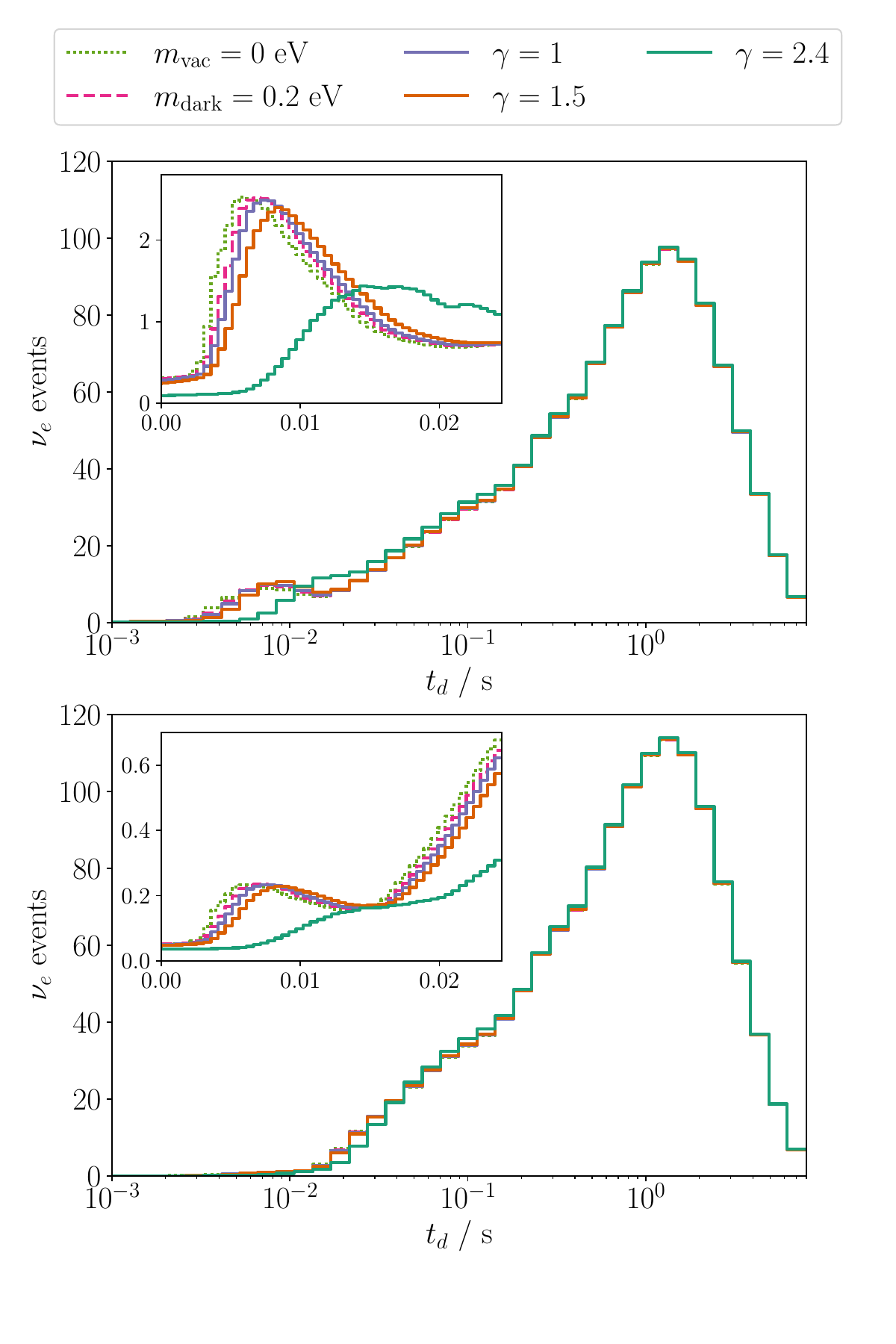}
\vspace{-1cm}
    \caption{Energy-integrated time distributions of the detected $\nu_e$ events in DUNE for IO (top) and NO (bottom). Dotted lines correspond to $m_{\rm vac} = 0$ eV, whereas all other curves assume $m_{\rm dark} = 0.2$ eV. Dashed lines show the expected $\nu_e$ rate without a DM spike, whereas solid lines illustrate the effect of crossing a spike region along the neutrino path, for different $\gamma$. The inset zooms into the neutronization phase.}
    \label{fig:ev_rates_DUNE}
\end{figure}

Fig.~\ref{fig:ev_rates_DUNE} represents the time distribution of the detected $\nu_e$ events in DUNE, integrated over energy and assuming $m_{\rm vac} = 0$ eV (green dotted line) and $m_{\rm dark} = 0.2$ eV in both inverted ordering (IO, top panel) and normal ordering (NO, bottom panel). Our choice of $m_{\rm dark}$ has been taken considering the expected projected bound from KATRIN \cite{KATRIN:2021dfa,KATRINCollaborationKATRINCollaboration2005_270060419}.
The total number of expected events in each scenario is given by $R \equiv \int R(t,E)\,dt\,dE$, where $t \equiv t_d \in [0, 9]$ s.

The effect due to the presence of a DM halo, in the absence of a spike, along the neutrino path is considered (magenta dashed line), as well as the one induced by crossing DM spike regions (solid lines), evaluated for different spike density normalizations $\gamma$. 
The presence of DM halos shifts the expected SN neutrino flux to later times, resulting in a spread-out detection rate at late times, especially for lower energy neutrinos. 
This effect is further enhanced in presence of DM spike regions.
As expected, the effect is more distinct for the neutronization phase (zoom boxes) as compared to the other phases, where the emission time dominates over the delay.

\noindent\textbf{\emph{Statistical analysis -- }}Following the method outlined in \cite{Pompa:2022cxc}, we evaluate the sensitivity of DUNE to $m_{\rm dark}$ and $\gamma$ for both NO and IO.
We consider scenarios where neutrinos either cross or avoid a DM spike region around the GC, in order to quantify the impact of the latter on the mass sensitivity. 
To this end, we define a symmetric DM spike region around the GC 
to avoid the singularity of the cusped NFW profile along the GC direction. We fixed $R_{\rm spike} = 0.21$ kpc and vary $\gamma$. We have verified that within the range of interest of $\gamma$, the change in $R_{\rm spike}$ is less than $\mathcal{O}(20)\%$. This would not affect our study. 
We generate toy datasets for DUNE, consisting of detected times $\delta t_i$ and energies $E_i$ of events from a SN at $D=10$ kpc along a trajectory defined by $\varphi$, following Ref.~\cite{Pompa:2022cxc}.

Each dataset is fitted with three free parameters: the refractive neutrino mass $m_{\rm dark}$ and spike normalization $\gamma$, which enter the rate via the delay $\Delta t_{\rm dark}$, and a time offset $t_{\rm off}$, defined as the difference between the Earth arrival of the first SN neutrino and the detection of the first event ($k=1$). 
The fitted emission time for event $k$ is given by $t_{k,\rm fit} = \delta t_k - \Delta t_{{\rm dark},k}(m_{\rm dark},\gamma) + t_{\rm off}$.
To evaluate the DUNE sensitivity projections, we adopt an un-binned likelihood method,  $\mathcal{L}_i(m_{\rm dark},t_\text{off}, \gamma) = e^{-R_i}\prod_{k=1}^{R_i}\int R(t_k,E_k)G_k(E)dE~$~\cite{Pagliaroli:2010ik},
with $G_k$ accounting for a Gaussian energy smearing. 

To assess the DUNE's sensitivity to $m_{\rm dark}$, the $\chi^2$ function, $\chi_i^2(m_{\rm dark}, t_\text{off}, \gamma) = -2 \log(\mathcal{L}_i(m_{\rm dark},t_\text{off}, \gamma))~$, is minimized with respect to both $t_\text{off}$ and $\gamma$ and then combined with the results from all toy datasets for a given scenario to account for statistical fluctuations.
To evaluate the maximal possible impact of a DM spike, we fix $\gamma=2.4$, corresponding to 
the steepest spike profile considered in the literature, and minimize only with respect to 
$t_{\rm off}$. The same analysis procedure is applied to evaluate DUNE's sensitivity to $\gamma$, 
this time minimizing both $t_{\rm off}$ and $m_{\rm dark}$. Since we assume that $m_{\rm dark}$ accounts for a dominant portion of the neutrino mass, we do not allow $m_{\rm dark}\rightarrow0$, being this unphysical.

\begin{figure}[!t]
    \centering
    \includegraphics[width=\columnwidth]{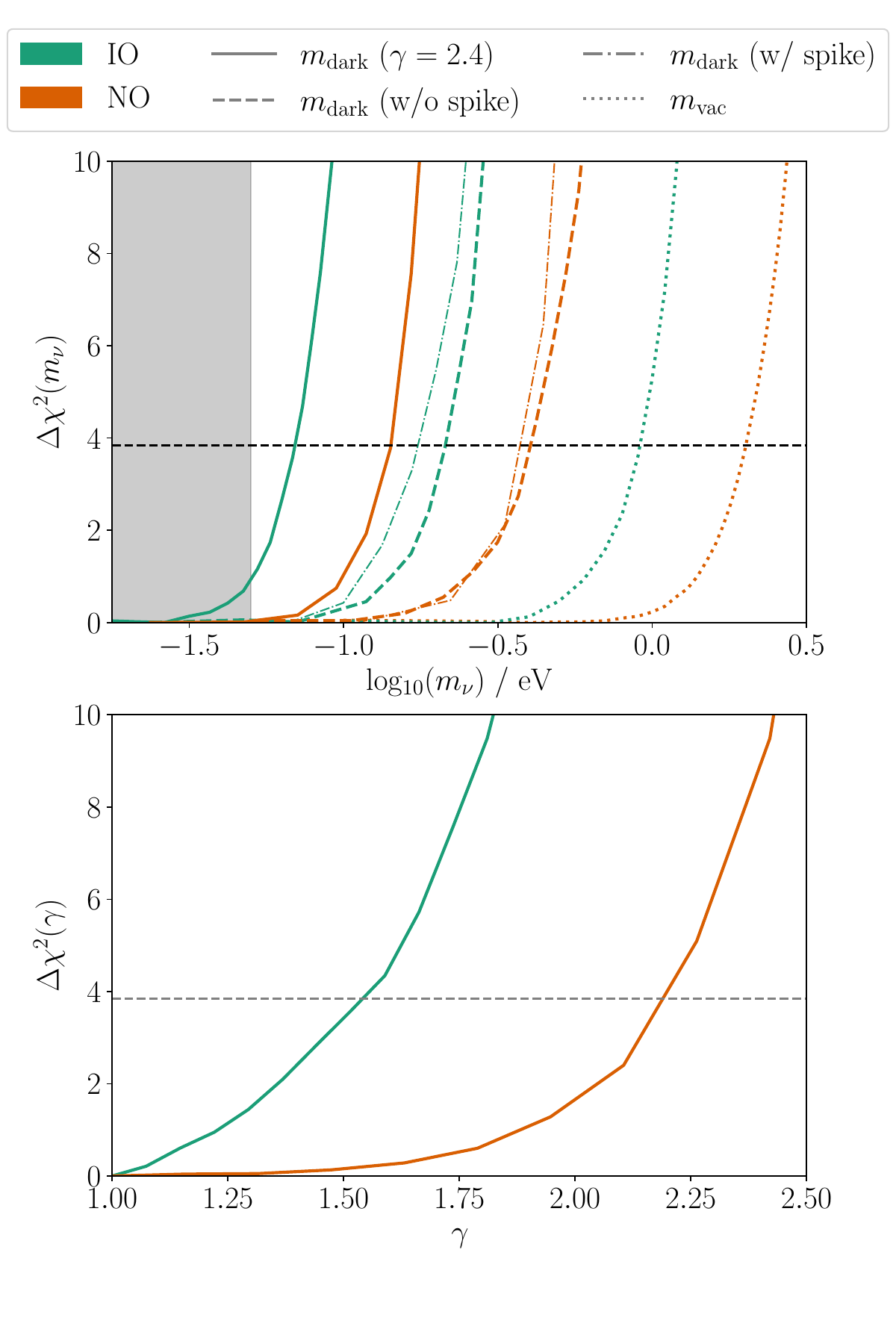}
    \vspace{-1cm}
    \caption{$\Delta\chi^2$ profiles for DUNE generated samples, assuming a source at $D = 10$~kpc.
    The horizontal dashed lines mark the $95\%$~C.L. bound.
    (Top): $\Delta\chi^2$ as a function of $\log_{10}(m_\nu)$ after marginalizing over $t_{\rm off}$. Dotted lines give $m_\nu \equiv m_{\rm vac}$ from \cite{Pompa:2022cxc}. Solid and dashed lines represent $m_\nu \equiv m_{\rm dark}$, where the neutrino crosses or avoids the DM spike region, respectively. Dot-dashed lines refer to the sensitivity after marginalizing over $\gamma$ too. Shaded area marks the parameter space ruled out by oscillation constraints.
    (Bottom): \dchigamma as a function of $\gamma$ after marginalizing over both $t_{\rm off}$ and $m_{\rm dark}$}. 
    \label{fig:Dchi2_profiles}
\end{figure}

\noindent \textbf{\emph{Results -- }}
For each mass ordering, we consider two scenarios: neutrinos propagating without entering the DM spike region at the GC, and neutrinos crossing the spike.

The resulting DUNE upper bound sensitivities on $m_{\rm dark}$ are shown in Fig.~\ref{fig:Dchi2_profiles} (top), and summarized in Tab.~\ref{tab:upper_bounds}.
The horizontal dashed line marks the $95\%$ C.L., corresponding to $\Delta \chi^2(m_{\rm dark}) = 3.84$.
For reference, the dotted lines reproduce the projected DUNE sensitivity to $m_{\rm vac}$ from \cite{Pompa:2022cxc}. Our analysis has been carried out by fixing the neutrino oscillation parameters to their best-fit values \cite{Esteban:2024eli,deSalas:2020pgw,Capozzi:2018ubv}, while the other uncertainties follow from~\cite{Pompa:2022cxc}.

\begin{table}[!h]
      \centering
\begin{tabular}{c @{\hskip 0.5cm} c @{\hskip 0.5cm} c @{\hskip 0.5cm} c @{\hskip 0.5cm} c}
\toprule
Ordering & $m_{\rm vac}$ / eV & \multicolumn{2}{c}{\mdark / eV} & $\gamma$ \\
\midrule
 & & w/o spike & w/ spike & \\
\midrule
IO   & $0.91$   & $0.21$   & $0.07~(0.17)$   & $1.54$ \\
\midrule
NO   & $2.01$  & $0.40$   & $0.14~(0.37)$   & $2.19$ \\
\bottomrule 
\end{tabular}
\caption{Upper bounds at $95\%$ C.L. on neutrino mass in both vacuum ($m_{\rm vac}$) and refractive mass (\mdark) scenarios, as well as on the DM spike normalization ($\gamma$). For cases including the spike region, the table reports the most stringent bound obtained for a fixed $\gamma = 2.4$, along with the bound obtained after marginalizing over $\gamma$ (in parenthesis, referring to the dot-dashed lines in Fig.~\ref{fig:Dchi2_profiles}).} 
\label{tab:upper_bounds}
\end{table}

The solid and dashed lines represent cases in which neutrinos traverse or avoid the DM spike, 
respectively. For the spike, we adopt $\gamma=2.4$, corresponding to one of the steepest 
spike profiles considered in the literature, thereby providing an estimate of the maximal 
enhancement.  
For other spike slopes, the sensitivity curves lie between the dashed and solid lines and are omitted for clarity. 
The dot-dashed lines refer to the sensitivity after marginalizing over $\gamma$ too. As expected, marginalization over $\gamma$ reduces the sensitivity to the upper bound on $m_{\rm dark}$.

The same analysis procedure is applied to estimate the sensitivity to the spike normalization 
$\gamma$.
The results are presented in Fig.~\ref{fig:Dchi2_profiles} (bottom), and summarized in Tab.~\ref{tab:upper_bounds}. 
These bounds quantify DUNE’s potential to probe the steepness of the DM spike profile, independently 
of the absolute refractive mass scale. The sensitivity to $\gamma$ depends on both the 
mass ordering and the assumed spike contribution, with steeper spikes yielding 
more pronounced time-delay enhancements. While uncertainties remain in the theoretical modeling of 
spike formation and survival, these results indicate that neutrino observations from a galactic 
SN could provide a novel and direct probe of DM density profiles in the innermost regions 
of the Milky Way. Taken together, the complementary sensitivities to $m_{\rm dark}$ and $\gamma$ 
highlight the power of DUNE SN observations to simultaneously test neutrino properties and 
the galactic DM distributions.

Note that vacuum and refractive masses induce parametrically distinct time delays: $\Delta t_{\rm vac}$ depends only on the source distance, whereas $\Delta t_{\rm dark}$ depends on the $\rho_{\phi}(r)$ LOS integral, and therefore on the angle with respect to the GC. 
The two effects coincide only along a fine-tuned curve in the $(m_{\rm dark},\gamma)$ plane, which remains distinguishable once the DUNE timing resolution $\delta t\sim\mathcal{O}(1~\mu\mathrm{s})$ is included.

\noindent\textbf{\emph{Conclusions -- }}In this work, we have investigated the potential of the DUNE experiment to probe refractive neutrino masses arising from interactions with DM, with particular attention to the impact of DM spikes near the GC. Using time-of-flight delays of neutrinos from a core-collapse 
SN at 10~kpc, we performed a detailed statistical study to evaluate DUNE’s sensitivity to both the refractive neutrino mass, $m_{\rm dark}$, and the spike normalization parameter, $\gamma$.

Our results show that the presence of a DM spike can substantially enhance the sensitivity of 
DUNE to $m_{\rm dark}$, improving projected bounds by upto a factor of three compared to smooth-halo expectations. In the most optimistic scenarios, DUNE could probe $m_{\rm dark} \sim \mathcal{O}(0.1)~$eV, a regime well below current SN neutrino time-of-flight mass limits, and complementary to terrestrial oscillation and kinematic measurements.
Furthermore, we demonstrated that DUNE observations could place competitive bounds on the spike normalization $\gamma$, offering a novel way to probe the distribution of DM near the GC. This establishes SN neutrino timing as a novel and competitive method for testing both the particle nature of neutrinos and the DM distribution in the innermost regions of our galaxy.

Some caveats remain. The modeling of DM spikes is subject to significant astrophysical uncertainties. Likewise, our analysis neglected collective neutrino oscillations and Earth-matter effects. Furthermore, neutrino-ULDM interactions can also deflect the neutrinos coming towards the Earth, causing an additional time delay~\cite{Visinelli:2024wyw}.
Taken together, these results highlight the unique complementarity of astrophysical neutrino observations and laboratory searches in the study of neutrino properties.

\textbf{\emph{Acknowledgments -- }}The authors acknowledge support of the Institut Henri Poincaré (UAR 839 CNRS-Sorbonne Université), and LabEx CARMIN (ANR-10-LABX-59-01). MS would like to thank Alexei Smirnov for useful discussions on the coherence time scales of ultralight dark matter. MS also acknowledges support from the Early Career Research Grant by Anusandhan National Research Foundation (project number ANRF/ECRG/2024/000522/PMS).

\bibliography{biblio}

\clearpage
\onecolumngrid

\begin{center}
    {\Large\bfseries Supplemental Material}
\end{center}

\vspace{1cm}

\twocolumngrid

\noindent\textbf{\emph{Complete sensitivity study -- }} 
A full likelihood analysis of the event rate in a DUNE-like detector allows to study how the sensitivity on the presence of a dark matter (DM) spike region depends on both the DM-induced neutrino mass, $m_{\rm dark}$, and the DM spike normalization parameter, $\gamma$. 

\begin{figure}[t]
    \centering
    \includegraphics[width=0.9\linewidth]{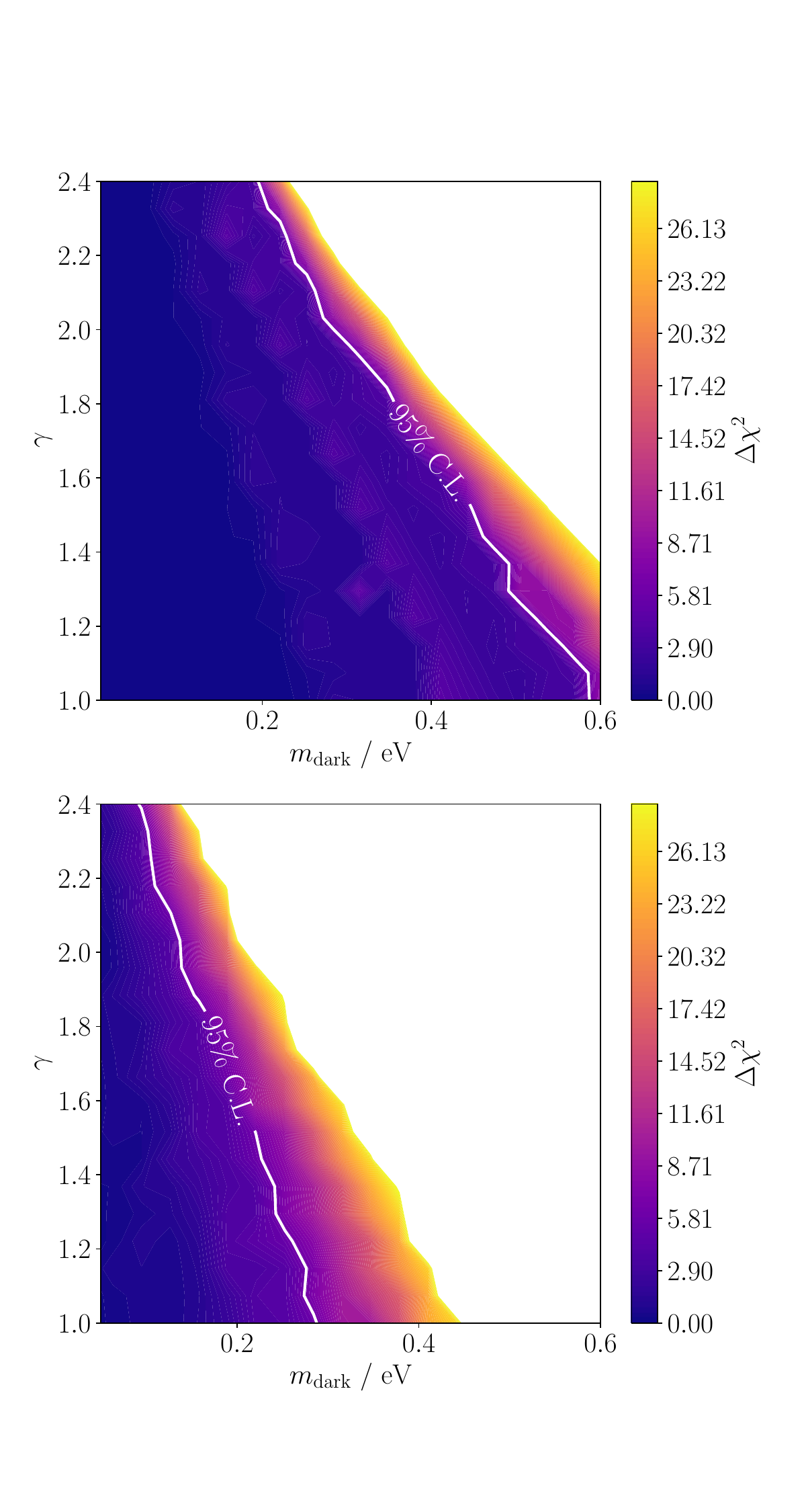}
    \caption{2D sensitivity contours in the $m_{\rm dark} - \gamma$ plane for normal ordering (top) and inverted one (bottom). The white lines correspond to the $95\%$ C.L., and the white areas mark the parameter region above the $5\sigma$ significance level. The contours have been evaluated over a single DUNE-like event toy generation, and agree with the results presented in the main analysis.}
    \label{fig:2D_likelihood}
\end{figure}

By adopting the same likelihood function $\mathcal{L}(m_{\rm dark}, t_{\rm off}, \gamma)$ defined in the main manuscript, and marginalising it over the detector time offset, $t_{\rm off}$, we obtain the 2D contours illustrated in Fig.~\ref{fig:2D_likelihood}, showing the sensitivity dependence on $m_{\rm dark}$ and $\gamma$ for both the normal ordering (top) and inverted ordering (bottom) cases.
The white lines mark the $95\%$ C.L. sensitivity in the $m_{\rm dark} - \gamma$ contour plane, whose upper limit is set at the $5\sigma$ level. 
Higher sensitivities lie in the white areas.
These contours refer to only one representative case, corresponding to a single SN neutrino event toy generation in DUNE, and agree with the results presented in Tab.~\ref{tab:upper_bounds} once averaged over multiple event generations.\\

\noindent\textbf{\emph{Distinguishing vacuum and refractive neutrino mass scenarios -- }} 
To address whether a DM–induced refractive mass can mimic the effect of a vacuum mass, we perform a statistical analysis, taking into account the information carried by the full event statistics expected in a DUNE-like detector setup.
\begin{figure}[b!]
    \centering
\includegraphics[width=0.85\linewidth]{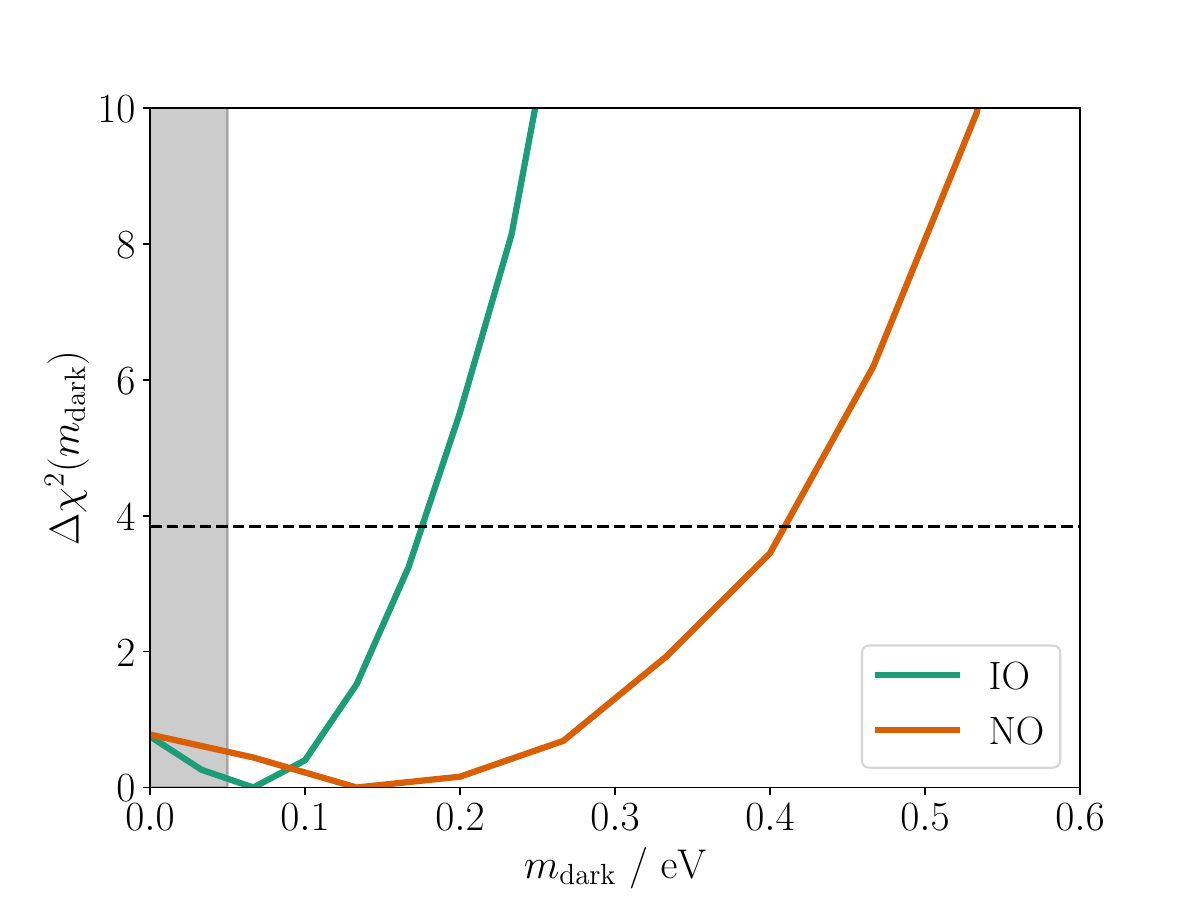}
\caption{$\Delta\chi^2(m_{\rm dark})$ profiles for both IO and NO, obtained by marginalising the likelihood over both $t_{\rm off}$ and $\gamma$.
    The minima correspond to the $m_{\rm dark}$ values that would reproduce the same time delay induced by $m_{\rm vac} = 0.2$ eV in each ordering. 
    The shaded area marks the parameter space region ruled out by oscillation data.}
    \label{fig:sensitivity}
\end{figure}

We adopt the same statistical approach described in the main manuscript, defining the same likelihood function $\mathcal{L}(m_{\rm dark}, t_{\rm off}, \gamma)$ and marginalising it over both the detector time offset, $t_{\rm off}$, and DM spike normalization, $\gamma$.
Instead of assuming massless neutrinos as the null hypothesis, we generated, this time, a toy DUNE dataset by imposing $m_{\rm vac} = 0.2$ eV (projected KATRIN sensitivity, sampled from the dashed magenta distribution of Fig.~\ref{fig:ev_rates_DUNE}), and evaluated the DUNE sensitivity to $m_{\rm dark}$.
Doing so, we test the hypothesis that neutrinos have a dark mass that, in the presence of a DM spike, will induce a finite time delay, which might mimic that induced by a finite vacuum mass.

\begin{figure}[b!]
    \centering
\includegraphics[width=0.95\linewidth]{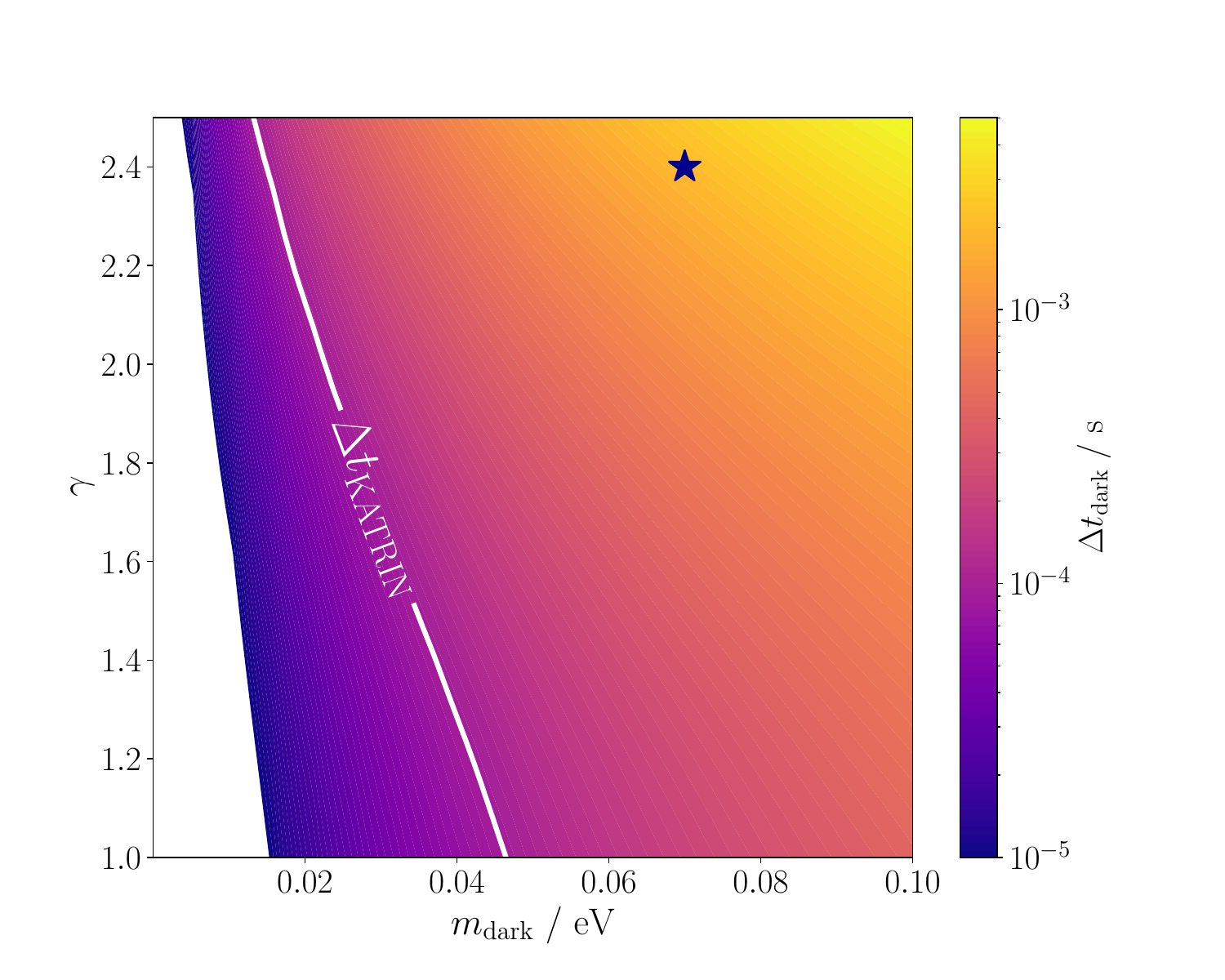}
    \caption{Contour plot of the refractive time delay $\Delta t_{\rm dark}$ in the $(m_{\rm dark},\gamma)$ plane,
assuming a representative neutrino energy of $E=15$~MeV and a supernova distance of $D=10$~kpc.
For comparison, the white curve corresponds to the vacuum-induced delay
$\Delta t_{\rm vac}$ under the same assumptions, taking $m_{\rm vac}=0.2$~eV.
The blue star indicates the point $\star \equiv  (m_{\rm dark},\gamma)=(0.07~\mathrm{eV},\,2.4)$, which yields the strongest bound obtained in our analysis and reported in Tab.~\ref{tab:upper_bounds}.  The shaded white region denotes parameter values for which the induced delay falls below the DUNE timing resolution $\delta t$, and is therefore experimentally inaccessible.}
    \label{fig:dt_contours}
\end{figure}
\begin{figure}[t!]
    \includegraphics[width=0.9\linewidth]{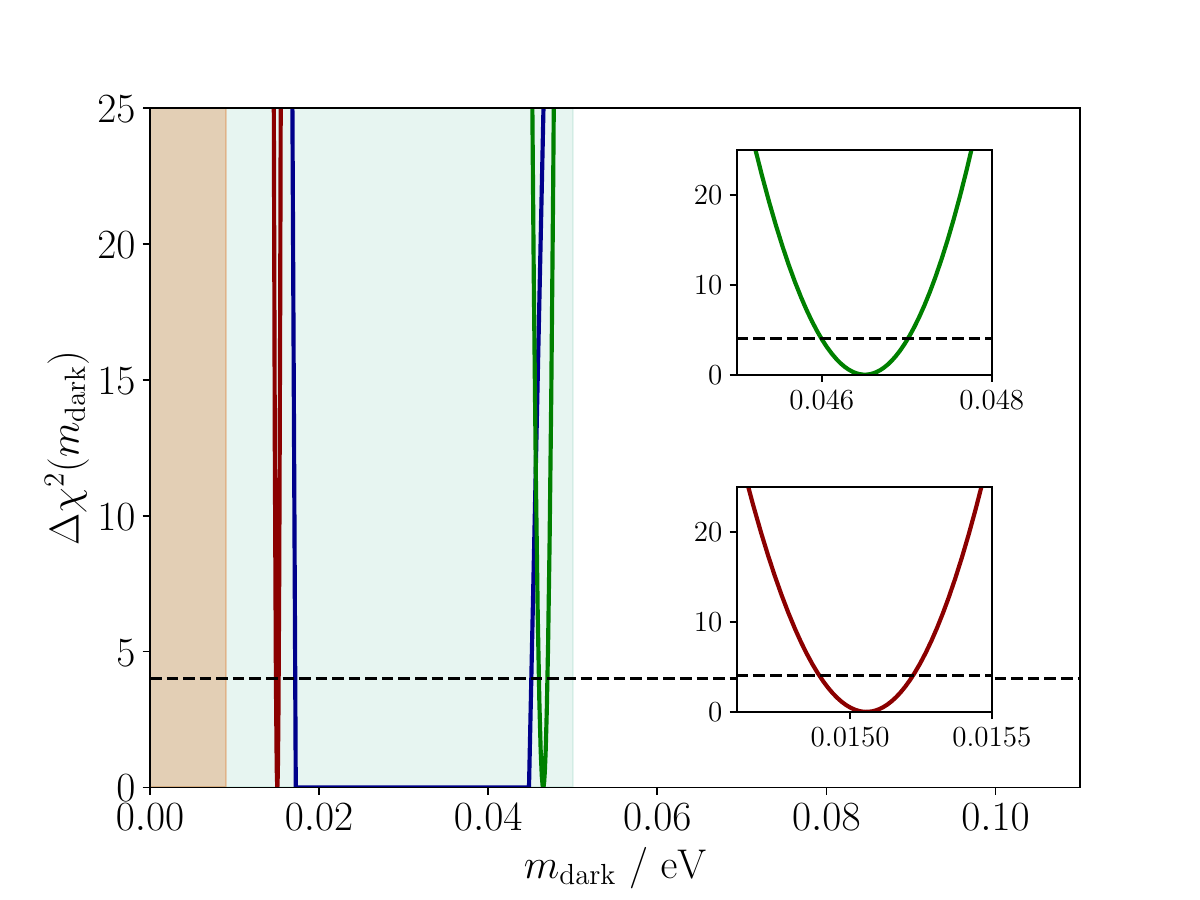}
    \caption{Sensitivity along the degeneracy line of Fig.~\ref{fig:dt_contours}, defined by the time delay induced by $m_{\rm vac} = 0.2$ eV. 
    Red (green) line corresponds to the $\Delta\chi^2$ profile of Eq.~\ref{eq:chi2_discrimination} when $\gamma=2.4$ ($\gamma=1$). 
    The minimum indicates the $m_{\rm dark}$ value that would induce the same time delay of $m_{\rm vac} = 0.2$ eV, which lies along the line. 
    When keeping $\gamma$ free, after marginalising over it along the degeneracy, all the sensitivity is lost in the whole $m_{\rm dark}$ mass range.
    Shadowed green (orange) area marks the parameter space excluded by oscillation data in IO (NO).}
    \label{fig:sensitivity_along_degeneracy}
\end{figure}

The result of this analysis is illustrated in Fig.~\ref{fig:sensitivity}.
We find that from an event statistics, DUNE will not be able to distinguish at $95\%$ C.L. between the time delay induced by the two scenarios, as long as the $m_{\rm dark}\lesssim 0.17\,{\rm eV} \,\,\textcolor{black}{(0.41\,{\rm eV} )}$ in IO (NO).
This is primarily because we are convoluting the time of arrival in the event rate computation, and hence it is possible for a dark mass with an appropriately chosen spike parameter, $\gamma$, to mimic the time delay induced by a finite vacuum mass. This is expected because the refractive mass scenario is chosen to mimic the effects of a finite vacuum mass, so it is possible that in the presence of a spike, a smaller refractive mass can induce the same time delay as a vacuum mass of $m_{\rm vac}=0.2\,$eV. 

The upper bounds are comparable with the $m_{\rm dark}\simeq 0.17\,{\rm eV} \,\,\textcolor{black}{(0.37\,{\rm eV} )}$ obtained for IO (NO) in the main analysis, and reported in parenthesis in Tab.~\ref{tab:upper_bounds}, showing no changes in the DUNE sensitivity when assuming $m_{\rm vac} \neq 0$.
This information also emerges from Fig.~\ref{fig:ev_rates_DUNE}: given the almost identical event rates expected to be detected when $m_{\rm vac} = 0$ and $m_{\rm vac} = 0.2$ eV (corresponding to the dotted green and dashed magenta lines, respectively), it is possible for DUNE to not distinguish between the two scenarios.

However, if DUNE were capable of directly measuring time delays, the analysis could be restricted to the comparison of the corresponding vacuum or dark time-of-flight delay induced within the DUNE time resolution. 

Figure~\ref{fig:dt_contours} shows the contours of the refractive time delay, $\Delta t_{\rm dark}$, in the $(m_{\rm dark},\gamma)$ plane for a benchmark supernova at $D=10~\text{kpc}$ and neutrino energy $E=15~\text{MeV}$. 
The blue star represents the point of the parameter space $\star \equiv (m_{\rm dark},\gamma)=(0.07~\mathrm{eV},\,2.4)$ corresponding to the strongest limit obtained in the main analysis and reported in Tab.~\ref{tab:upper_bounds}.
On the same plot, we overlay the vacuum-induced time delay, $\Delta t_{\rm vac} = 9.15 \times 10^{-5}$ s, corresponding to a neutrino mass $m_{\rm vac}=0.2~\text{eV}$, shown as a white reference curve. 
Points lying on this curve represent combinations of $(m_{\rm dark},\gamma)$ that give the same total time delay as a $0.2$~eV vacuum mass. 
Only along this fine-tuned line can the two scenarios be degenerate at the level of a single time-delay measurement.

To quantify this, we perform a likelihood analysis by defining
\begin{equation}
    \chi^{2}(m_{\rm dark}, \gamma) 
    = \frac{\big(\Delta t_{\rm dark}(m_{\rm dark},\gamma)-\Delta t_{\rm vac}(0.2~\text{eV})\big)^{2}}
    {\delta t^{2}}\,,
    \label{eq:chi2_discrimination}
\end{equation}
where $\delta t\sim\mathcal{O}(1~\mu\mathrm{s})$ is the DUNE timing resolution.
The resulting one-dimensional profile likelihood is shown in Fig.~\ref{fig:sensitivity_along_degeneracy}.
Here, red and green lines, zoomed in the two insets, refer to the cases in which $\gamma$ in Eq.~\ref{eq:chi2_discrimination} has been fixed at the values of 2.4 and 1, respectively, corresponding to the boundary of the $\gamma$ range considered in this analysis.
The minimum of each curve corresponds to the value of $m_{\rm dark}$ that would reproduce the same time delay induced by a vacuum mass $m_{\rm vac}=0.2$~eV for the given DM spike normalisation, and belongs to the degeneracy curve of Fig.~\ref{fig:dt_contours}.
Horizontal dashed lines define the DUNE's $2\sigma$ sensitivity.

This result shows that, even for small departures from the fine-tuned degeneracy line and when relying on direct time-delay measurements,  DUNE's excellent time resolution would allow for discrimination between vacuum and refractive neutrino mass scenarios at the level of $2\sigma$ or better if $\gamma$ is known, within the all phenomenologically accessible parameter space. 
This, of course, requires a direct measurement of the time delay induced in the two scenarios, which is difficult to obtain.
When marginalising over $\gamma$, the sensitivity is lost along the whole $m_{\rm dark}$ range to which the degeneracy belongs, as indicated by the blue line in Fig.~\ref{fig:sensitivity_along_degeneracy}.
Note, however, that this mass range belongs to the parameter space excluded by oscillation data, depicted as shadowed green (orange) area in IO (NO).

\end{document}